\documentclass[preprint,12pt,authoryear]{elsarticle}
\usepackage{color}

\usepackage{amssymb}

\journal{Journal of High Energy Astrophysics}

\begin{document}

\begin{frontmatter}



\title{Time Dependent Leptonic Modeling of Fermi II Processes in the Jets of Flat 
Spectrum Radio Quasars}


\author[OU]{C. Diltz}
\author[NWU,OU]{M. B\"ottcher}

\address[OU]{Astrophysical Institute, Department of Physics and Astronomy,
Ohio University, Athens, OH 45701, USA}
\address[NWU]{Centre for Space Research, North-West University, Potchefstroom,
2520, South Africa}

\begin{abstract}
In this paper, we discuss the light-curve features of various flaring scenarios in a time-dependent 
leptonic model for low-frequency-peaked blazars. The quasar 3C273 is used as an illustrative example. 
Our code takes into account Fermi-II acceleration and all relevant electron cooling
terms, including the external radiation fields generally found to be important in the modeling 
of the SEDs of FSRQs, as well as synchrotron self absorption and $\gamma\gamma$ pair-production. 
General parameters are constrained through a fit to the average spectral energy
distribution (SED) of the blazar by numerically solving the time-dependent Fokker-Planck 
equation for the electron evolution in a steady-state situation. We then apply perturbations 
to several input parameters (magnetic field, particle injection luminosity, acceleration
time scale) to simulate flaring events and compute time-dependent SEDs and light curves in 
representative energy bands (radio, optical, X-rays, $\gamma$-rays). Time lags between 
different bands are evaluated using a discrete cross correlation analysis. We find that 
Fermi-II acceleration has a significant effect on the distributions and that flaring events 
caused by increased acceleration efficiency of the Fermi II process will produce a correlation 
between the radio, optical and $\gamma$-ray bandpasses, but an anti-correlation between these 
three bandpasses and the X-ray band, with the X-rays lagging behind the variations in other
bands by up to several hours.
\end{abstract}

\begin{keyword}
Active and peculiar galaxies and related systems \sep $\gamma$-rays \sep Radiative transfer
\sep Elementary particle processes 

98.54.Cm \sep 95.85.Pw \sep 95.30.Jx \sep 95.30.Cq


\end{keyword}

\end{frontmatter}






\section{Introduction}
\label{intro}

Blazars represent a class of radio-loud Active Galactic Nuclei that consists of BL Lac objects 
and Flat Spectrum Radio Quasars (FSRQs). The spectral energy distributions (SED) of blazars is 
characterized by two broadband, nonthermal components that span from the radio to UV or X-ray 
wavelengths and from x-rays to high-energy $\gamma$-rays. The extreme inferred isotropic-equivalent 
$\gamma$-ray luminosities, combined with rapid variability in different bandpasses, in some cases, 
down to just a few minutes, provides evidence for strong Doppler boosting in these sources. This 
is considered to be the result of beamed emission from relativistic jets closely alisgned with our
line of sight. It is generally accepted that the low-energy spectral component is synchrotron
emission of relativistic electrons/positrons. For the origin of the high-energy SED component, 
two different approaches have been discussed, referred to as leptonic and hadronic models
\citep[for a review of both types of models, see, e.g.,][]{boettcher07,Boettcher12}. In 
the leptonic scenario, the X-ray to $\gamma$-ray emission is due to the inverse 
Compton scattering off the relativistic electrons, with the target photon fields either being 
the synchrotron photons within the emission region (SSC = synchrotron self Compton), or 
photons external to the jet (EC = external Compton). The external photon fields can include the accretion
disk \citep{Dermer92,Dermer93}, the broad line region (BLR), \citep{Sikora94,Blandford95}, 
or even an infra-red emitting dust torus (IR) that surrounds the central accretion flow onto 
supermassive black hole \citep{Blazejowski00}. Leptonic models are widely used and have been 
relatively successful in modeling the SEDs and some variability features of blazars. In hadronic 
models \citep[e.g.,][]{mb92,Mastichiadis95,Muecke01,Muecke03,Mastichiadis05,Boettcher13}, 
$\gamma$-rays are the result of proton synchrotron radiation as well as $\pi^0$-decay and
synchrotron and Compton radiation from secondary particles in photo-pion induced cascades,
presuming the existence of ultrarelativistic protons in the emission region. While such 
models have also had success in modeling the SEDs of blazars and remain viable, rapid
variability observed in blazars is more readily explained in terms of the much shorter
acceleration and cooling time scales of relativistic leptons. Therefore, in this work, 
we will focus on leptonic models.\\

The shapes of the spectral components provide insight into the underlying particle distribution 
that is producing the emission. Simple power-law and broken power-law electron distributions with
parameters chosen ad-hoc, have often been invoked in order to model the SEDs of blazars. Alternatively,
log-parabolic electron distributions have been successfully employed to produce the curved synchrotron
and Compton spectra observed in many blazars \citep{Massaro04,Massaro06,Cerruti13,Dermer14}. The 
log-parabola function is characterized by two variables that describe the spectral parameter of the 
electron distribution and the spectral curvature of the distribution. The log parabolic shape has 
been shown to be analytically related to a stochastic acceleration mechanism, in which the acceleration 
probability decreases with energy \citep{Rani11,Massaro06}. Such a connection of log-parabolic
spectra and acceleration mechanisms naturally arises in solutions of the time dependent Fokker-Planck 
equation that contains a momentum diffusion term, indicative of Fermi II acceleration, when
the evolution reaches equilibrium \citep{Tramacere11,Massaro06}. They showed that the spectral 
curvature is inversely proportional to the momentum diffusion coefficient, since the diffusion term 
acts to broaden the shape of the particle distribution. \\

Second order Fermi acceleration is therefore a viable mechanism for producing log-parabola particle 
spectra which may be hard enough to reproduce the hard spectra of $\gamma$-ray emission observed in 
several TeV blazars \citep{Lefa11,Asano13}. It has also been shown that relativistic Maxwellian 
electron distributions can result from stochastic acceleration processes balanced by radiative 
losses \citep{1Schlickeiser84}. For the full time dependent Fokker-Planck equation incorporating 
Fermi II acceleration, general solutions have been found using Green's functions and the application 
of spectral operators \citep{Stawarz08,Tramacere11}. Solutions to the Fokker-Planck equation incorporating
both Fermi I and Fermi II processes, have been developed for the application of the transport of energetic 
ions \citep{Becker06}. Solutions have also been obtained that consider both Fermi I and Fermi II 
acceleration, and radiative losses in the Thomson regime \citep{1Schlickeiser84,2Schlickeiser84}. 
However, when Klein-Nishina effects on the electron cooling rates, as well as absorption processes
in the radiation transfer problem, are taken into account, one needs to resort to numerical
solutions of the Fokker-Planck equation. \cite{Asano13} developed a time dependent Leptonic model 
that incorporated Fermi II processes to study the hard spectrum of the blazars Mrk 421 and 1ES 1101-232.
The curvature of the electron spectrum, as well as the hard $\gamma$-ray spectra could be reproduced 
by a model that utilizes a stochastic momentum diffusion process (Fermi II). \\

In this paper, we use a time-dependent Leptonic model that incorporates Fermi acceleration and self-consistent
radiative losses, including synchrotron and Compton scattering on internal (SSC) and external (EC) radiation
fields as well as synchrotron self-absorption and $\gamma\gamma$ absorption and pair production. 
The purpose of this paper is to investigate the effects of various flaring scenarios, including
Fermi-II acceleration, in external-Compton dominated blazars to complement the study for SSC-dominated
sources by \cite{Asano13}. Therefore, while our code is applicable to all types of blazars, we here
focus on its application to FSRQs. We describe
the model and underlying assumptions in Section \ref{theory}. We use our code to study the influence of 
Fermi-II acceleration on the quasi-equilibrium particle distribution and light-curve features, including 
possible time delays beween variations in different frequency bands. These features are studied with 
parameters motivated by an SED fit to the FSRQ 3C 273, described in Section \ref{spectrum}. Once we have 
obtained appropriate baseline parameters, we choose a set of input parameters (specifically, the particle 
injection luminosity, the magnetic field, and the acceleration time scale) to perturb them in the form 
of a Gaussian in time, in order to study the light curves in the radio, optical, x-ray and $\gamma$-ray 
bandpasses (Section \ref{lightcurve}). In Section \ref{correlation}, we perform a discrete correlation 
function analysis on the light curves obtained in the preceding section, to determine possible time lags 
between the selected bandpasses. We summarize and discuss our results in Section \ref{results}. 
Throughout this paper, a cosmology with $\Omega_{m} = 0.3$, $\Omega_{\Lambda} = 0.7$, and 
$H_{0} = 70 km s^{-1} Mpc^{-1}$ is used.

\section{\label{theory}Model Setup}

Our model is based on a single, homogenous emission region of radial size $R$ which moves relativistically
with bulk Lorentz factor $\Gamma$ along a pre-existing jet structure, oriented at a small angle $\theta_{\rm obs}$
with respect to our line of sight. Throughout the paper, unprimed quantities denote values in the co-moving 
frame of the emission region, while primed quantities denote values in the stationary AGN frame. The emission 
region is pervaded by a homogeneous, randomly oriented magnetic field of strength B. The size of the emission 
region is constrained by the observed variability time scale, $\Delta t$, through

\begin{equation}
R = \frac{c \cdot \Delta t \cdot \delta}{1 + z}
\label{Rconstraint}
\end{equation}	 

\noindent where $z$ is the redshift to the source and $\delta = \left( \Gamma [ 1 - \beta_{\Gamma}
\cos\theta_{\rm obs}] \right)^{-1}$ is the Doppler factor. 
 
A population of an ultra-relativistic electrons is continuously injected. We assume that the electron 
injection spectrum is in the form of a power-law distribution with the functional form

\begin{equation}
Q(\gamma, t) = Q_{0} (t) \gamma^{-q} H(\gamma; \gamma_{min}, \gamma_{max})
\label{injectionspectrum}
\end{equation}

\noindent where $H(\gamma; \gamma_{min}, \gamma_{min})$ denotes the Heaviside function 
defined by $H = 1$ if $\gamma_{min} \le \gamma \le \gamma_{max}$, and $H = 0$ otherwise. 
The normalization factor for the injection spectrum is determined through the injection 
luminosity by

\begin{equation}
Q_{0} = 
\left\{
\begin{array}{ll}
	\frac{L_{inj} (t)}{V_{b} m_{e} c^{2}} \frac{2-q}{\gamma_{2}^{2-q} - \gamma_{1}^{2-q}} & \mbox{if $q \ne 2$}, \\
	\frac{L_{inj} (t)}{V_{b} m_{e} c^{2} ln(\frac{\gamma_{2}}{\gamma_{1}})} & \mbox{if $q = 2$}.
\end{array}
\right.
\label{Q0}
\end{equation}					

\noindent where $V_{b}$ denotes the comoving blob of the emission region and $m_{e}$ denotes 
the rest mass of an electron. 

The time evolution of the electron distribution is found by numerically solving the time-dependent 
Fokker-Planck equation, which is given in the following form:

\begin{equation}
\frac{\partial n_{e} (\gamma, t)}{\partial t} = \frac{\partial}{\partial \gamma}[\frac{1}{(a+2) 
\cdot t_{acc}} \cdot \gamma^{2} \cdot \frac{\partial n_{e} (\gamma, t)}{\partial \gamma}] - 
\frac{\partial}{\partial \gamma} (\dot{\gamma}_{rad} \cdot n_{e} (\gamma, t)) + Q(\gamma, t) 
- \frac{n_{e} (\gamma ,t)}{t_{esc}}
\label{FPequation}
\end{equation}

\noindent 
where $a = v_{s}^{2}/v_{A}^{2}$, $v_{A}$ represents the Alfven velocity, $v_{s}$ represents 
the shock velocity. In this study, a value of $a = 10^{-3}$ is chosen.
$\dot\gamma_{rad}$ denotes the radiative (synchrotron and Compton) losses, 
taking into account Klein-Nishina effects \citep[e.g.,][]{1Boettcher97}. Synchrotron 
losses are governed by the strength of the randomly oriented magnetic field within the 
emission region. Inverse Compton losses are governed by the scattering of the 
electrons with the synchrotron photons that they produce (SSC) or by the external radiation 
fields surrounding the black hole (EC). These radiation fields include 
emission directly from the accretion disk, emission reprocessed by the Broad Line Region 
(BLR), and radiation emitted by a dusty torus. 

In adition to radiative losses, the electron distribution is subjected gyro-resonant 
wave-particle interactions with hydromagnetic turbulence described by a turbulent plasma
wave spectrum $I(k) \propto k^{-p}$ with index $p$, where $k$ is the wave number of 
turbulent plasma waves. 

If the energy density of the plasma waves is small compared to the energy density 
of the magnetic field (quasi-linear approximation), then the diffusion coefficient 
becomes a power law function of the form

\begin{equation}
D(\gamma) = K \cdot \gamma^{p}
\label{Diffusioncoefficient}
\end{equation}	

\noindent In this work, we consider a diffusion coefficient with a spectral index $p = 2$ 
(hard sphere scattering). This makes the acceleration time scale independent of energy. 
The normalization of the diffusion coefficent is given by $K = 1/([a+2] t_{acc})$. 
The values $t_{acc}$ and $t_{esc}$ represent the acceleration 
and escape time scales, respectively. We parameterize the escape time scale in terms of
the light crossing time scale as $t_{esc} = \eta R/c$ where $\eta \ge 1$. 

The Fokker-Planck equation is solved through an implicit Crank-Nichelson scheme that 
converts the partial differential equation into a tri-diagonal set of linear equations. 
The solution to the linear equations is then found through a tri-diagonal matrix algorithm. 
This method has the advantage of being unconditionally stable, allowing us to use arbitrarily
large time steps to numerically solve the partial differential equation when approaching 
equilibrium. 

Simultaneously with the Fokker-Planck equation (\ref{FPequation}) for the electrons, 
we solve a separate evolution equation for the photon field in the emission region:

\begin{equation}
\frac{\partial n_{ph} (\nu, t)}{\partial t} = \frac{4 \pi}{h \nu} 
\cdot j_{\nu} (t) - n_{ph} (\nu, t) \cdot 
(\frac{1}{t_{\rm esc, ph}} + \frac{1}{t_{\rm abs}})
\label{photonevolution}
\end{equation}

\noindent where $j_{\nu}$ denotes the emissivity due to the various radiation mechsnisms,
$t_{\rm esc, ph} = 4R/3c$ is the photon escape time scale and $t_{\rm abs}$ denotes 
the absorption time scale due to synchrotron self absorption and gamma-gamma absorption. 
The absorption time scale can be defined through the opacities $\tau$, as

\begin{equation}
t_{\rm abs} = \frac{R}{c \cdot (\tau_{\rm SSA} + \tau_{\gamma \gamma})}
\label{tabs}
\end{equation}

\noindent where $\tau_{SSA}$ and $\tau_{\gamma \gamma}$ denote the synchrotron-self-absorption and 
$\gamma\gamma$ absorption opacities. With the solution to the photon field at any given time 
step, we then compute the emerging (observable) broadband spectrum in the observer's frame 
through

\begin{equation}
f_{\nu'}^{\prime} (t^{\prime}) \equiv \nu' F'_{\nu'} (t') = 
\frac{h \cdot \nu^{2} \cdot n_{ph} (\nu, t) \cdot \delta^{4} 
\cdot V_{co}}{4 \pi d_{L}^{2} \cdot t_{\rm esc, ph}}
\label{nuFnu}
\end{equation}

At every time step, separate subroutines are used to compute the emission coefficients for the
various radiation processes in the co-moving frame of the emission region. The synchrotron 
emission coefficient is evaluated as

\begin{equation}
j_{\nu, syn} (t) = \frac{1}{4 \pi}\int_{0}^{\infty} d\gamma n_{e} (\gamma, t) \cdot 
P_{\nu} (\gamma)
\label{jnusy}
\end{equation}	

\noindent where the term $P_{\nu} (\gamma)$ denotes the spectral synchrotron power of 
a single lepton. The spectral synchrotron power is approximated by \citep{Boettcher12}:

\begin{equation}
P_{\nu}(\gamma) = \frac{32\pi c}{9\Gamma(4/3)} \cdot r_{e}^{2}(\frac{m_{e}}{m})^{2} \cdot 
u_B \gamma^2 \cdot \frac{\nu^{1/3}}{\nu_{c}^{4/3}} e^{-\nu/\nu_{c}}
\label{Pnusy}
\end{equation}

\noindent where $u_{B}$ denotes the energy density of the magnetic field and $r_{e}$ denotes the classical 
electron radius. The critical frequency for the synchrotron spectrum is $\nu_{c} = 
(4.2 \times 10^{6} B(m_{e}/m)) \cdot \gamma^{2} Hz$. 

The SSC emissivity is calculated using the solution of \cite{Jones68} for Compton scattering
of an isotropic radiation field by an isotropic distribution of relativistic electrons. Our code uses 
the entire co-moving photon field as targets for Compton scattering, thus incorporating higher-order 
SSC scattering. 

For the external radiation field from the accretion disk, we assume that the disk is in the form of 
an Shakura-Sunyaev disk \citep{Shakura74} with the following intensity profile:

\begin{equation}
I_{\epsilon}^{\prime SS} (\Omega'; \tilde R') = \frac{3GM \dot{m}}{16 \pi^{2} R^{3}} \cdot \varphi(\tilde R') 
\cdot \delta (\epsilon^{\prime} - \frac{C}{\tilde R^{3/4}})
\label{ISS}
\end{equation}	 

\noindent where $\epsilon' = h\nu'/m_{e} c^{2}$ is the photon energy normalized to the 
electron rest energy in the AGN frame and $\varphi(\tilde R')$ is defined by

\begin{equation}
\varphi(\tilde R') = 1 - \beta_{i} \cdot (R_{i}^{\prime}/R')^{1/2}
\label{phi}
\end{equation}

\noindent where $\beta_{i}$ denotes the fraction of angular momentum captured by the black hole at the 
radius $R_{i}$, the innermost stable circular orbit around the black hole. The constant $C$ is defined 
by

\begin{equation}
C = 1.51\times 10^{-4} (\frac{l_{edd}}{\eta_{f} M_{9}})^{1/4}
\label{Cconstant}
\end{equation}	

\noindent and $\tilde R' = R'/R_{g}^{\prime}$, with $R_{g}^{\prime}$ denoting the gravitational 
radius. With this representation for the intensity, we can compute the observed $\nu F_{\nu}$ 
flux of the accretion disk \citep{Dermer09}:

\begin{equation}
f_{\epsilon', SS}^{\prime} = \frac{l_{edd} L_{edd}}{2 \pi d_{L}^{2} \eta_{f} \tilde R_{min}^{\prime}} 
\cdot (\frac{\epsilon'}{\epsilon_{max}^{\prime}})^{4/3} \cdot exp(-\epsilon'/\epsilon_{max}^{\prime})
\label{fSS}
\end{equation}

\noindent We next consider an isotropic, external blackbody radiation field of temperature $T_{BB}$
surrounding the emission region. This is an appropriate representation for a thermal IR radiation 
field from a dust torus (with $T_{bb} \lesssim 1000$~K), but also produces an external-Compton spectrum
in good agreement with that resulting from a full BLR radiation field for $T_{bb} \sim$ a few $10^3$~K
\citep{Boettcher13}, as long as the emission region is not located far beyond the outer boundary of
the BLR. The spectral energy density of the external radiation field is

\begin{equation}
u'(\epsilon') = K \frac{\epsilon^{\prime 3}}{exp(\epsilon'/\Theta) - 1}
\label{uepsilon}
\end{equation}	

\noindent where $K$ denotes the normalization constant and $\Theta = kT / (m_e c^2)$ denotes the 
dimensionless temperature parameter. The normalization constant is constrained equating 
$\int_0^{\infty} u' (\epsilon') \, d\epsilon'$ to the expected energy density of the radiation 
field in the AGN frame

\begin{equation}
u_{ext}^{\prime} = \frac{L_{d}^{\prime} \tau}{4 \pi R_{ext}^{\prime 2} c}
\label{uext}
\end{equation}

\noindent where $L_{d}$ denotes the total luminosity of the disk and $\tau$ denotes the fraction of 
the disk's radiation that's reprocessed by either the dust torus or the BLR and reemitted as thermal 
radiation, and $R_{ext}$ denotes the radius of the (assumed spherical) reprocessing material. 
In order to evaluate the emission coefficients for EC scattering in the comoving frame, we need to
transform the spectral energy density (Equ. \ref{uepsilon}) from the AGN frame to the co-moving
frame through \citep{Dermer09}:

\begin{equation}
u(\epsilon, \Omega) = \frac{u'(\epsilon', \Omega')}{\Gamma^{3} (1+\beta\mu)^{3}}
\label{uprime}
\end{equation}

\noindent where $\mu$ denotes the cosine of the angle between the normalized, negative bulk 
velocity of the emission region and direction of propagation of photons in the co-moving frame,
and $u'(\epsilon', \Omega') = u'(\epsilon')/(4\pi)$ under the assumption of isotropy in the AGN rest
frame. The resulting spectral energy density in the comoving frame is then used to evaluate the 
emission coefficient in the comoving frame. 

For time dependent modeling, evaluating the full expressions for the EC emission coefficients,
involving full integrations over the solid angles of the target photon field and the electron
distributions in the comoving frame, is impractically time-consuming. Therefore, in order to
save computing time, we assume that photons from both the accretion disk and the isotropic 
radiation field are boosted into the relativistic blob in the forward direction, thus replacing
the angular characteristic of Equation (\ref{uprime}) by a $\delta$ function $\delta(\mu + 1)$. 
For the accretion disk, we invoke the near field approximation \citep{Dermer09}, which is valid
as long as the relativistic blob is near the black hole:

\begin{equation}
u_{disk} (\epsilon, \Omega) = \frac{u_{NF}}{2 \pi} \cdot 
\delta(\epsilon - \Gamma \epsilon^{\prime}_{*}/2) \cdot \delta(\mu + 1)
\label{uprimedisk}
\end{equation}

\noindent where $u_{NF}^{\prime}$ denotes the total energy density of the accretion disk in the 
near-field regime in the comoving frame and $\epsilon^{\prime}_{*}$ denotes the peak energy of 
the accretion-disk emission. 

For the isotropic radiation field, we construct the energy density in the comoving frame using

\begin{equation}
u_{ext} (\epsilon, \Omega) = \frac{15 u_{IR}}{2 \pi^{5} 
(\Theta)^{4}} \cdot \frac{\epsilon^{3}}{\exp(\epsilon/\Theta) - 1} 
\cdot \delta(\mu + 1)
\label{uprimethermal}
\end{equation}

\noindent With equations (\ref{uprimedisk}) and (\ref{uprimethermal}), we use the relations given 
in \cite{Dermer09} to compute the corresponding EC emission coefficicents. 

Once we have the combined photon field of all the radiation fields in the comoving frame, we 
compute the $\gamma\gamma$ absorption opacity \citep{Dermer09}, and the pair production rate
\citep{2Boettcher97}. The produced pair spectrum is added to the solution of the Fokker-Planck
equation from the current time step, and $\tau_{\gamma\gamma}$ is included in the calculation
of the photon absorption time scale (Equ. \ref{tabs}) for the next time step.

\section{\label{spectrum}Steady State Spectrum}

The purpose of this paper is a generic study of variability features caused by variations of 
individual emission-region parameters in the model described in the previous section. To choose
realistic baseline parameters for this study, we perform a fit to the time-averaged SED of the 
FSRQ 3C273 \citep[data taken from][]{abdo10}, based on an equilibrium solution obtained with our
code with time-independent input parameters. The equilibrium model is fully determined through 
the following list of input parameters: The magnetic field $B$, the observed variability time 
scale, $\Delta t_{\rm var}$, the bulk Lorentz factor $\Gamma$, the observing angle $\theta_{\rm obs}$, 
the low- and high-energy cutoffs $\gamma_{\rm min, max}$ and spectral index $q$ of the electron 
injection spectrum, the electron injection luminosity $L_{\rm inj}$, the accretion disk luminosity
$L'_{\rm disk}$, the initial distance of the emission region from the central black hole, $R_{\rm axis}$,
the characteristic extent $R'_{\rm ext}$, energy density $u'_{\rm ext}$, and blackbody temperature
$T_{\rm BB}$ of the external radiation field, and the ratio between the acceleration and escape 
time scales, $t_{\rm acc} / t_{\rm esc}$. 

Several of these parameters may be either directly measured or constrained through observations. 
Specifically, for 3C273, we have the following observables \citep[see][for references to the 
observational data]{Boettcher13}: $z = 0.158$, $\beta_{\perp, \rm app} = 13$ (the 
apparent transverse velocity of individual jet components, normalized to the speed of light), 
$\Delta t_{\rm var} \sim 1$~d, $L_{\rm disk} = 1.3 \times 10^{47}$~erg~s$^{-1}$, and $L_{\rm BLR} 
= 9.1 \times 10^{45}$~erg~s$^{-1}$. The observed apparent superluminal speed implies a limit to the 
bulk Lorentz factor of $\Gamma > 13$. We choose the observing angle as $\theta_{\rm obs} = 1/\Gamma$
so that $\delta = \Gamma$, and the size of the emission region is then constrained by Equ.
(\ref{Rconstraint}). In the SED of 3C273, the accretion disk component is directly visible
as a prominent Big Blue Bump, which facilitates reliable estimates of the black hole mass 
and Eddington ratio, $l_{\rm Edd} = L_{\rm disk} / L_{\rm Edd}$. The observed BLR luminosity 
is related to the disk luminosity through $L_{\rm BLR} = \tau \, L_{\rm disk}$, and these 
quantities are related to the energy densities of the respective radiation fields in the 
co-moving frame of the emission region through \citep{Dermer09}

\begin{equation}
\frac{u_{ext}}{u_{B}} = \frac{8 \cdot L'_{d} \cdot \tau \cdot \Gamma^{2}}{3 \cdot B^{2} 
\cdot R_{ext}^{\prime 2} \cdot c}
\label{uext_uB}
\end{equation}

\begin{equation}
\frac{u_{disk}}{u_{B}} = \frac{6 \cdot G \cdot M_{bh} \cdot \dot{m} \cdot 
\Gamma^{2} \cdot 0.023}{B^{2} \cdot c \cdot R_{axis}^{3}}
\label{udisk_UB}
\end{equation}

We may relate the observed synchrotron peak frequency to the Doppler factor, magnetic field and
peak electron energy, assuming that the peak energy corresponds to $\gamma_{\rm min}$. 
Assuming that SSC scattering is in the Thomson regime, the peak location of the SSC power will 
then be located at $\nu_{SSC} = \nu_{syn} \cdot \gamma_{\rm min}^2$. Given a constraint on
the Doppler factor, this provides estimates for $\gamma_{\rm min}$ and $B$. Unfortunately,
the spectral slope of the synchrotron spectrum is difficult to constrain for 3C 273 due to the 
substantial contribution from the accretion disk in the optical regime, which is masking much of 
the synchrotron emission. 

\begin{figure}[t]
\begin{center}
\includegraphics[height=0.45\textheight]{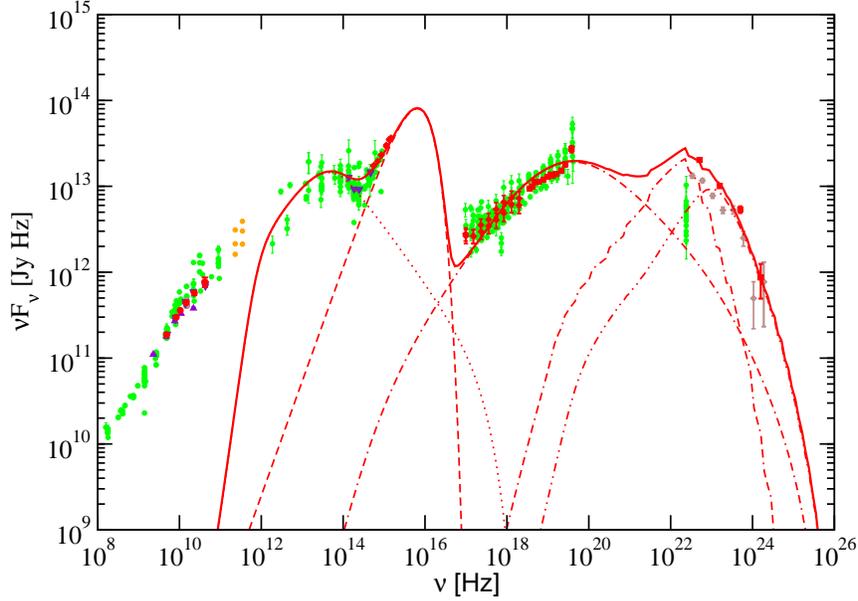}
\caption{Equilibrium fit to the time-averaged SED of 3C 273. See Table \ref{parameters} for
parameter values. The line styles denote: solid = overall fit, dotted = synchrotron, dashed = 
accretion disk, single-dot-dashed = SSC, dot-double-dashed = EC (accretion disk), double-dot-dashed 
= EC (isotropic radiation field).}
\label{SEDfit}
\end{center}
\end{figure}

Within the framework of the observational constraints, the remaining parameters are varied to 
obtain an acceptable fit to the SED of 3C273. Figure \ref{SEDfit} shows the SED fit obtained, 
with parameters listed in Table \ref{parameters}. Our fitting procedure is a "fit by eye" method. 
Due to the considerable number of adjustable parameters not constrained by observations, a 
detailed $\chi^{2}$ minimization procedure is infeasible. While our fit parameters might 
provide a reasonable estimate of the actual physical conditions in the emission region, 
the lack of a rigorous $\chi^{2}$ minimization procedure makes an error analysis impractiable. 
However, since the goal of this paper is the study of light curve features resulting from 
individual parameter variations, the exact value of any individual parameter is irrelevant 
for our purpose. 

\begin{table}
\centering
\begin{tabular}{cc}
\hline
Parameter & Value \\
\hline
$ B $ & $1.75$~G \\ 
$ R $ & $7.25 \times 10^{15}$~cm \\
$\eta$ & 18.0 \\
$\Gamma$ & 14 \\
$\theta_{\rm obs}$ & $7.14 \times 10^{-2}$~rad \\
$ \gamma_{min} $ & $ 6.25 \times 10^{2} $ \\
$ \gamma_{max} $ & $ 1.0 \times 10^{5} $ \\
$q$ & 3.4 \\
$L_{\rm inj}$ & $5.6 \times 10^{42}$~erg~s$^{-1}$ \\
$M_{\rm BH}$ & $2.6 \times 10^{9} \, M_{\odot}$ \\
$l_{\rm Edd}$ & 0.395 \\
$R_{\rm axis}$ & 0.07~pc \\
$R'_{\rm ext}$ & 0.73~pc \\
$u'_{\rm ext}$ & $6.5 \times 10^{-5}$~erg~cm$^{-3}$ \\
$T_{\rm BB}$ & 6000~K \\
$ t_{acc}/t_{esc} $ & $1.5 \times 10^{-3} $ \\
\hline
\end{tabular}
\caption[]{\label{parameters}Parameter values used for the equilibrium fit to the SED of
3C273 (see Figure \ref{SEDfit}).}
\end{table}

Figure 1 illustrates that the SED of 3C273 is reproduced quite well with our fit. The optical to near UV 
radiation is fitted well by a combination of synchrotron and direct accretion disk emission. The value 
of $B = 1.75$~g is consistent typical values (of the order of 1 -- a few G) found in the modeling of 
FSRQs by other authors \citep[e.g.,][]{Ghisellini11,Boettcher13,Dermer14}. The X-ray spectrum is
fitted with a synchrotron self Compton component, and the {\it Fermi}-LAT data points are fitted with 
external-Compton radiation, also in agreement with most other modeling works which utilize external 
Compton scattering to reproduce the $\gamma$-ray emission of low-frequency-peaked blazars, as opposed 
to high frequency BL Lacs that can usually be well represented by pure synchrotron-self-Compton models.
In our fit, the external radiation fiels is a combination of radiation from the accretion disk and 
an isotropic external radiation field (representative of the BLR), as suggested by \cite{Finke10} 
to reproduce the spectral break in the {\it Fermi}-LAT spectrum of the FSRQ 3C~454.3. The choice of 
parameters for the isotropic radiation field is consistent with being related to the BLR, given the 
distance of the emission region from the black hole, $\sim 10^{17} cm$, the radial extent of the 
external field, $\sim 10^{18} cm$, and the blackbody temperature of $6 \times 10^{3} K$ \citep[see,
e.g.,][]{Boettcher13}. The distance from the black hole is also consistent with the near-field 
approximation adopted for the accretion disk radiation field. The radio emission is suppressed 
due to synchrotron self absorption, which suggests that the extended radio emission is likely 
due to synchrotron emission of electrons further down the jet.  

Our fit employs a moderate diffusive acceleration time scale of $t_{\rm acc} = 6.5 \times 10^3$~s, 
which is longer than the radiative cooling time scale of electrons at $\gamma_{\rm max}$. Therefore, 
the influence of Fermi-II acceleration on the presented steady-state fit is negligible. However, as 
we will see in the next sections, this is no longer the case for the flaring scenarios that we 
investigate.

\begin{figure}[t]
\begin{center}
\includegraphics[height=0.45\textheight]{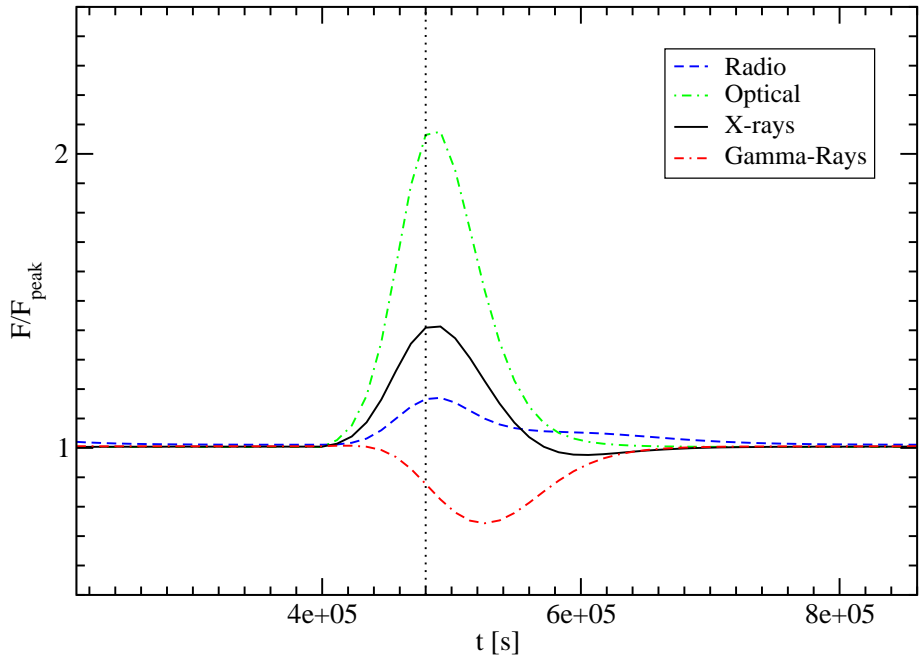}
\caption{Normalized lightcurves in the radio, optical, X-ray, and $\gamma$-ray bandpasses resulting
from the magnetic field perturbation (Equ. \ref{Bmodification}). The dotted vertical line indicates
the value of $t_0$. }
\label{B_lightcurves}
\end{center}
\end{figure}

\section{\label{lightcurve}Simulated Lightcurves}

\begin{figure}[t]
\begin{center}
\includegraphics[height=0.45\textheight]{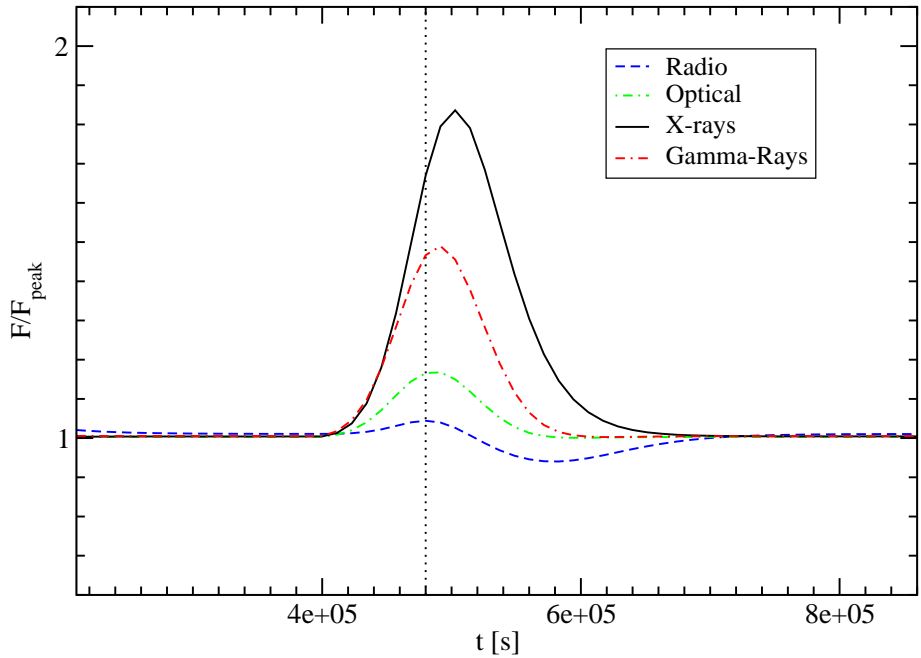}
\caption{Normalized lightcurves in the radio, optical, X-ray, and $\gamma$-ray bandpasses resulting 
from the injection luminosity perturbation (Equ. \ref{Lmodification}). The dotted vertical line 
indicates the value of $t_0$. }
\label{L_lightcurves}
\end{center}
\end{figure}

\begin{figure}[t]
\begin{center}
\includegraphics[height=0.45\textheight]{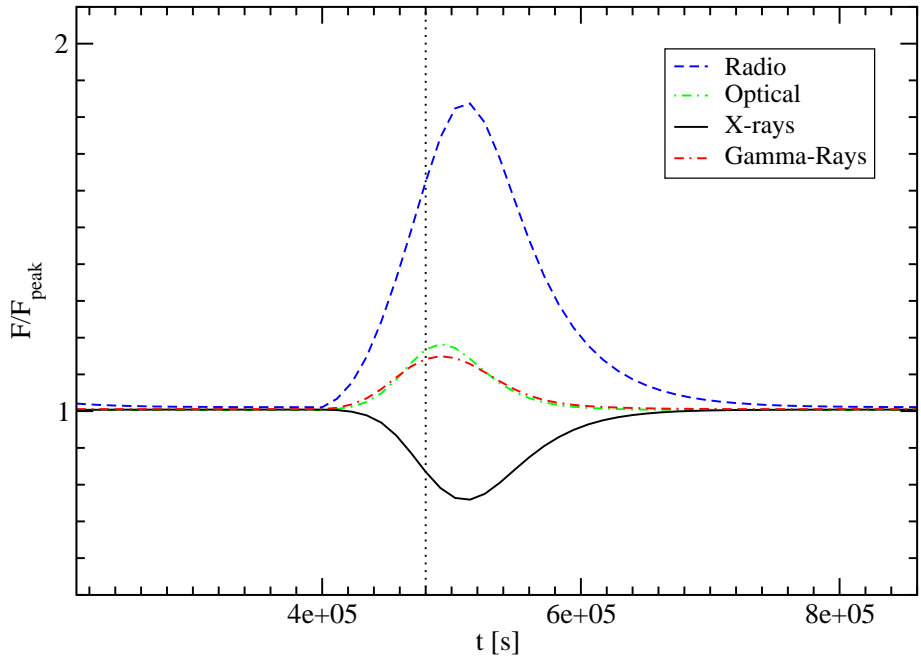}
\caption{Normalized lightcurves in the radio, optical, X-ray, and $\gamma$-ray bandpasses resulting 
from the acceleration time scale perturbation (Equ. \ref{taccmodification}). The dotted vertical 
line indicates the value of $t_0$. }
\label{t_lightcurves}
\end{center}
\end{figure}

Starting with the steady state model setup for 3C 273 as described above, we now investigate the influence
of fluctuations of individual parameters on the time-dependent radiative output. After the simulation has 
reached equilibrium, one of the input parameters ($B$, $L_{\rm inj}$, or $t_{\rm acc}$) is modified in the 
form of a Gaussian perturbation in time. From the simulation outputs, we extract light curves in the radio, 
optical (R-band), X-ray and GeV $\gamma$-ray ({\it Fermi}-LAT) bandpasses. Specifically, We setup the time 
evolution of the magnetic field perturbation as

\begin{equation}
B(t) = B_0 + K_{B} \cdot e^{-(t - t_{0})^{2}/2\sigma^{2}}
\label{Bmodification}
\end{equation}

\noindent where $B_0 = 1.75$~G is the equilibrium value for the magnetic field, $K_{B} = 2$~G parameterizes 
the amplitude of the perturbation, and $t_{0}$ and $\sigma$ specify the time when the perturbation reaches 
its maximum and the characteristic time scale of the perturbation, respectively. The chosen perturbation 
for the injection luminosity has the same functional form,

\begin{equation}
L_{\rm inj} (t) = L_{\rm inj, 0} + K_{L} \cdot e^{-(t - t_{0})^{2}/2\sigma^{2}}
\label{Lmodification}
\end{equation}
	 
\noindent where $L_{\rm inj, 0} = 5.6 \times 10^{42}$~erg~s$^{-1}$ is the equilibrium injection luminosity 
and $K_{L} = 4.8 \times 10^{42}$~erg~s$^{-1}$ is the amplitude of the perturbation. The perburbation of the 
acceleration time scale is chosen in such a way that the acceleration time scale decreases to a minimum 
during the peak of the perturbation. This is achieved with the following parameterization:

\begin{equation}
t_{\rm acc} (t) = \frac{t_{\rm acc, 0}}{1 + K_{t} \cdot e^{-(t - t_{0})^{2}/2\sigma^{2}}}
\label{taccmodification}
\end{equation}

\noindent where $t_{\rm acc, 0}$ is the equilibrium value of the acceleration time scale and $K_t = 17$ 
characterizes the amplitude of the perturbation. For all three perturbations, we choose a width of 
$\sigma = 4 \times 10^5$~s, and a peak time of $t_0 = 6.7 \times 10^6$~s, corresponding to approximately 
2 and 30 light-crossing time scales through the emission region, respectively, both in the co-moving 
frame. The light curves (normalized to the respective peak fluxes) are shown in figures \ref{B_lightcurves} 
to \ref{t_lightcurves}.

As expected, the increase in the magnetic field (Figure \ref{B_lightcurves}) causes an increase 
in the synchrotron flux at all energies (specifically, radio and optical for the case studied 
here). The associated increase in the synchrotron photon energy density also causes a flare 
in the SSC-dominated X-ray emission. At the same time, this leads to increased radiative 
cooling without a change of the external radiation fields and, hence, a dip in the $\gamma$-ray 
light curve. This dip is delayed with respect to the maxima in the synchrotron and SSC light 
curves by approximately the radiative cooling time scale of $\gamma$-ray emitting electrons. 

A temporarily increased injection luminosity (Figure \ref{L_lightcurves}) initially causes a 
flare in all bandpasses. However, we find a delayed decrease of the radio flux following an 
initial, small-amplitude flare. This is explained by an increased density of relatively low-energy 
electrons, responsible for synchrotron self-absorption at radio wavelengths, delayed by the 
required radiative cooling time scale for newly injected electrons to reach Lorentz factors 
of $\lesssim 100$. 

A decreasing acceleration time scale (Figure \ref{t_lightcurves}) leads to more efficient 
acceleration of relativistic particles to higher energies. The electron spectrum becomes 
harder and extends to higher energies during this perturbation, which shifts all spectral 
components to higher frequencies. This leads to flaring behaviour in the radio, optical, 
and $\gamma$-ray bands. The X-ray flux, however, corresponding to the low-frequency branch of the 
SSC emission component, decreases due to the shift of the SSC component to higher frequencies. The flare in 
the radio bandpass is particularly pronounced in this simulation. Keeping in mind that the radio emission is
in the optically-thick (to SSA) regime, this can be explained by low-energy electrons being accelerated to
higher energies, out of the energy range contributing to synchrotron self-absorption at radio wavelengths, 
thereby reducing the effective number density of electrons for SSA. Due to the much steeper frequency
dependence of the SSA opacity compared to the synchrotron emissivity, the net effect is an increase in 
the emanating synchrotron flux. 

The light curves shown in Figures \ref{B_lightcurves} to \ref{t_lightcurves} reveal noticeable delays 
between the light curve features in different frequency bands. Tables \ref{B_peaks} -- \ref{t_peaks} 
list the simulated equilibrium fluxes $f_0$, peak fluxes $f_{pk}$, FHWM and peak times $t_{pk}$ of the 
light curves in the four studied frequency bands for the three perturbations investigated here. The 
fluxes, $f_{0}$ and $f_{pk}$, are given in units of erg~cm$^{-2}$~s$^{-1}$, while the FWHM and peak 
time, $t_{pk}$, is given in units of seconds.  

\begin{table}
\centering
\begin{tabular}{ccccc}
\hline
& Radio & R-Band & X-rays & $\gamma$-rays \\
\hline
$f_{0}$  & $7.5 \times 10^{-16}$ & $2.68 \times 10^{-11}$ & $1.78 \times 10^{-10}$ & $6.85 \times 10^{-10}$  \\
$f_{pk}$  & $8.71 \times 10^{-16}$ & $5.53 \times 10^{-11}$ & $2.50 \times 10^{10}$ & $5.1 \times 10^{-10}$  \\
$FWHM$ & $8.36 \times 10^{4}$ & $7.31 \times 10^{4}$ & $7.80 \times 10^{4}$ & $9.57 \times 10^{4}$  \\
$t_{pk}$ & $4.86 \times 10^{5}$ & $4.87 \times 10^{5}$ & $4.87 \times 10^{5}$ & $5.26 \times 10^{5}$  \\
\hline
\end{tabular}
\caption{Simulated light curve parameters for the case of the magnetic field perturbation. Fluxes, $f_{0}$ 
and $f_{pk}$, are given in units of erg~cm$^{-2}$~s$^{-1}$, while the FWHM and peak time, $t_{pk}$, are given 
in units of seconds.}
\label{B_peaks}
\end{table}
	 
\begin{table}
\centering
\begin{tabular}{ccccc}
\hline
& Radio & R-Band & X-rays & $\gamma$-rays \\
\hline
$f_{0}$  & $7.5 \times 10^{-16}$ & $2.68 \times 10^{-11}$ & $1.78 \times 10^{-10}$ & $6.85 \times 10^{-10}$  \\
$f_{pk}$  & $7.03 \times 10^{-16}$ & $3.13 \times 10^{-11}$ & $3.26 \times 10^{10}$ & $1.01 \times 10^{-9}$  \\
$FWHM$ & $9.43 \times 10^{4}$ & $6.45 \times 10^{4}$ & $8.51 \times 10^{4}$ & $7.81 \times 10^{4}$  \\
$t_{pk}$ & $5.74 \times 10^{5}$ & $4.87 \times 10^{5}$ & $5.03 \times 10^{5}$ & $4.89 \times 10^{5}$  \\
\hline
\end{tabular}
\caption{Simulated light curve parameters for the case of the injection luminosity perturbation. Fluxes, $f_{0}$ 
and $f_{pk}$, are given in units of erg~cm$^{-2}$~s$^{-1}$, while the FWHM and peak time, $t_{pk}$, are given 
in units of seconds.}
\label{L_peaks}
\end{table}

\begin{table}
\centering
\begin{tabular}{ccccc}
\hline
& Radio & R-Band & X-rays & $\gamma$-rays \\
\hline
$f_{0}$  & $7.5 \times 10^{-16}$ & $2.68 \times 10^{-11}$ & $1.78 \times 10^{-10}$ & $6.85 \times 10^{-10}$  \\
$f_{pk}$  & $1.36 \times 10^{-15}$ & $3.16 \times 10^{-11}$ & $1.35 \times 10^{-10}$ & $7.86 \times 10^{-10}$  \\
$FWHM$ & $1.04 \times 10^{5}$ & $7.37 \times 10^{4}$ & $9.01 \times 10^{4}$ & $9.08 \times 10^{4}$  \\
$t_{0}$ & $5.11 \times 10^{5}$ & $4.87 \times 10^{5}$ & $5.10 \times 10^{5}$ & $4.90 \times 10^{5}$  \\
\hline
\end{tabular}
\caption{Simulated light curve parameters for the case of the acceleration timescale perturbation. Fluxes, $f_{0}$ 
and $f_{pk}$, are given in units of erg~cm$^{-2}$~s$^{-1}$, while the FWHM and peak time, $t_{pk}$, are given 
in units of seconds.}
\label{t_peaks}
\end{table}

\noindent The predicted anti-correlation between the X-ray fluxes and the radio, optical, and $\gamma$-ray
fluxes found for the case of the acceleration timescale perturbation, is a particularly interesting feature.
These correlations and anti correlations could represent a tell-tale signature of flaring activity caused
by a temporary increase of the efficiency of Fermi II acceleration in the emission region.

\begin{figure}[t]
\begin{center}
\includegraphics[height=0.25\textheight]{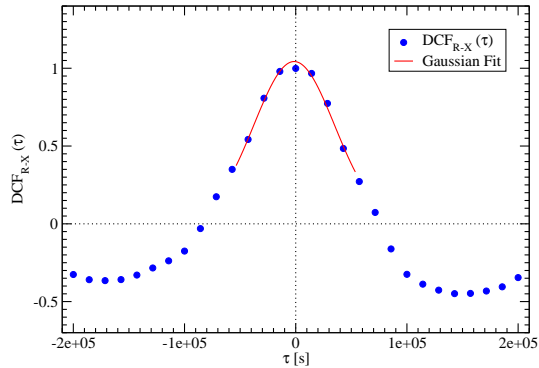}
\caption{Discrete correlation function between the optical (R-band) and X-ray bandpasses for the magnetic 
field perturbation case, along with a Gaussian fit to the DCF.}
\label{DCF_RXB}
\end{center}
\end{figure}

\begin{figure}[t]
\begin{center}
\includegraphics[height=0.30\textheight]{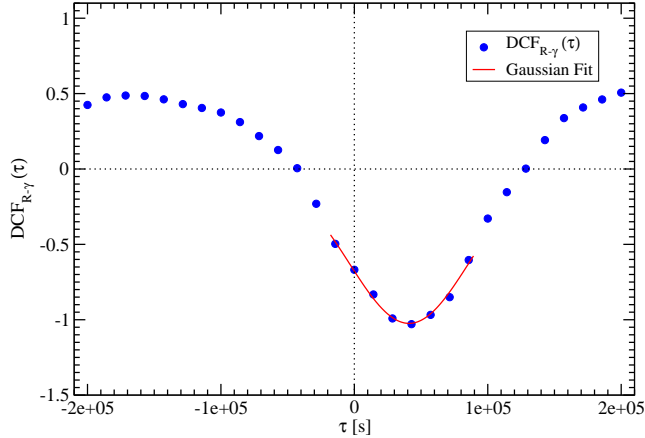}
\caption{Discrete correlation function between the optical (R-band) and $\gamma$-ray bandpasses 
for the magnetic field perturbation case, along with a Gaussian fit to the DCF. }
\label{DCF_RGB}
\end{center}
\end{figure}

\section{\label{correlation}Discrete Correlation Analysis of Lightcurve Bandpasses}

\begin{figure}[t]
\begin{center}
\includegraphics[height=0.30\textheight]{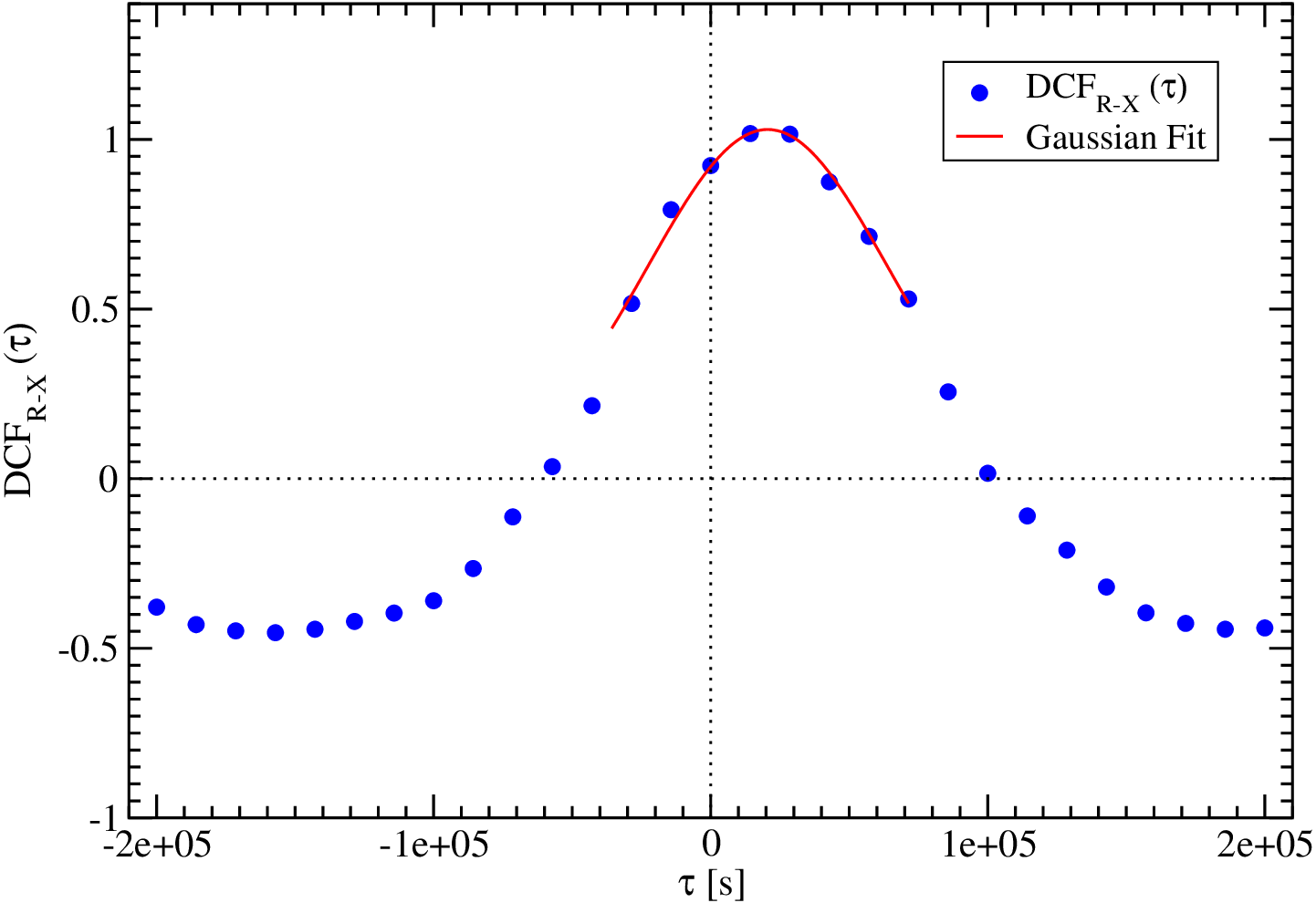}
\caption{Discrete correlation function between the optical (R-band) and X-ray bandpasses for the injection 
luminosity perturbation case, along with a Gaussian fit to the DCF.}
\label{DCF_RXL}
\end{center}
\end{figure}

\begin{figure}[t]
\begin{center}
\includegraphics[height=0.30\textheight]{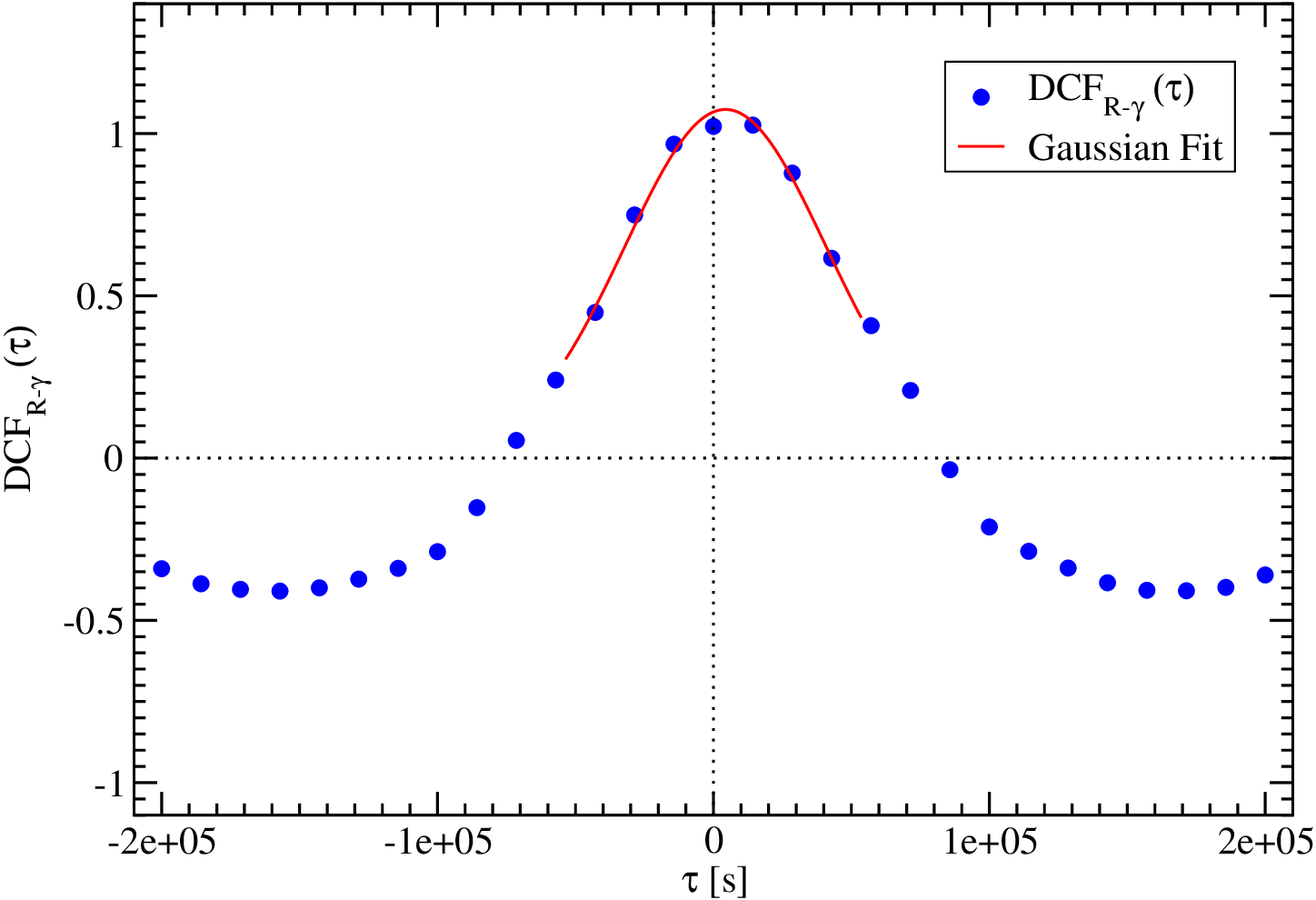}
\caption{Discrete correlation function between the optical (R-band) and $\gamma$-ray bandpasses for the  
injection luminosity perturbation case, along with a Gaussian fit to the DCF.}
\label{DCF_RGL}
\end{center}
\end{figure}

\begin{figure}[t]
\begin{center}
\includegraphics[height=0.30\textheight]{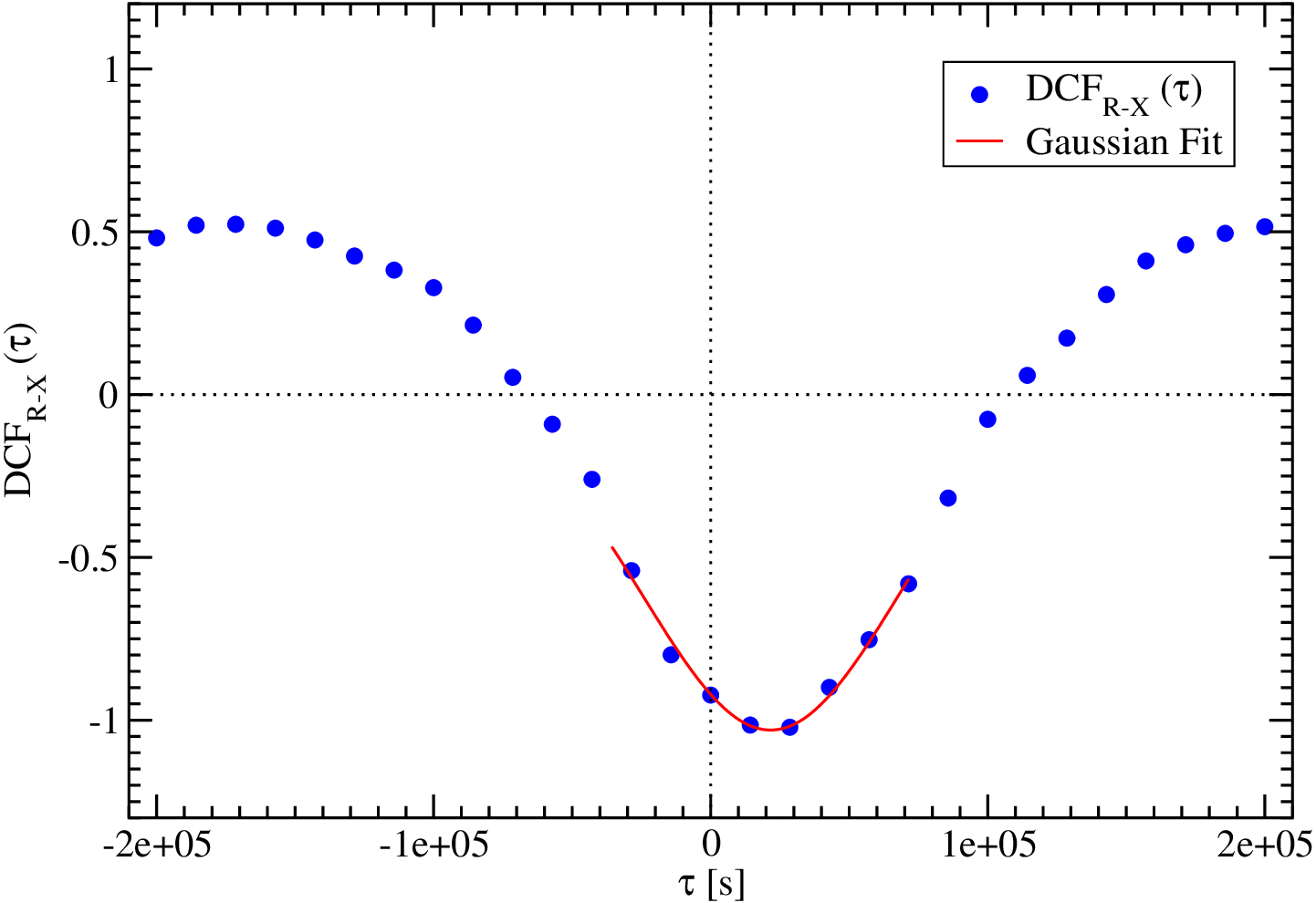}
\caption{Discrete correlation function between the optical (R-band) and X-ray bandpasses for the acceleration
time scale perturbation case, along with a Gaussian fit to the DCF.}
\label{DCF_RXt}
\end{center}
\end{figure}

\begin{figure}[t]
\begin{center}
\includegraphics[height=0.30\textheight]{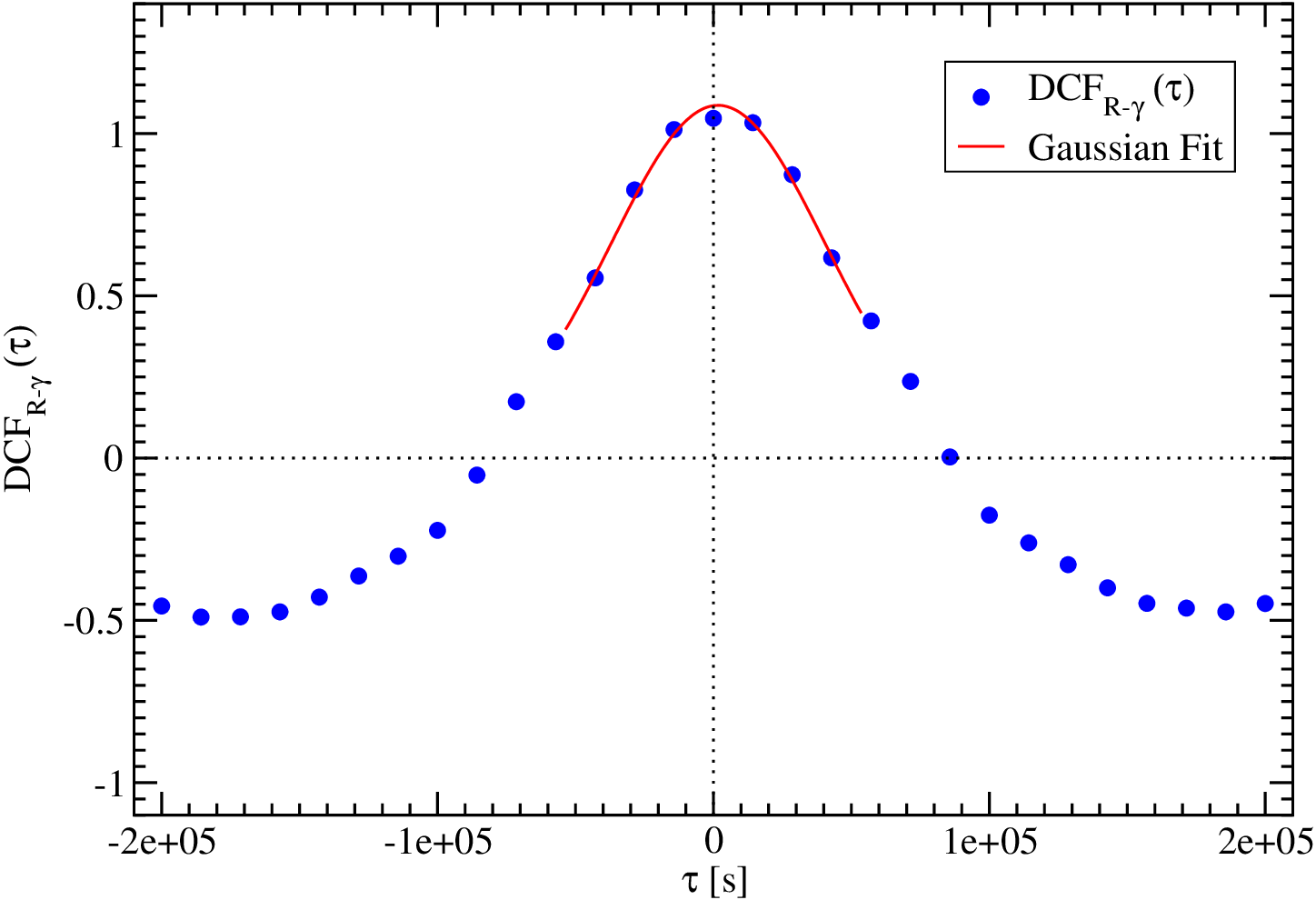}
\caption{Discrete correlation function between the optical (R-band) and $\gamma$-ray bandpasses for the 
acceleration time scale perturbation case, along with a Gaussian fit to the DCF.}
\label{DCF_RGt}
\end{center}
\end{figure}

In order to be able to directly compare our predictions to light curve features extracted from
observational data, we apply a discrete correlation function (DCF) analysis \citep{Edelson88} between 
the light curves at the various bandpasses investigated here, as is routinely done for data from 
flux-monitoring campaigns on blazars to study correlations/anticorrelation and time lags between 
different frequency bands. We arbitrarily assign a relative error of $1\%$ of the flux values to any 
simulated light curve point in order to be able to apply a $\chi^2$ minimization technique to fit a 
phenomenological Gaussian function to the obtained DCFs. For comparison with observational data, 
which typically have the most complete temporal light curve coverage in the optical and $\gamma$-ray ({\it Fermi}-LAT) --- and occasionally 
also X-ray --- bands, we here focus on the cross correlations between the optical and the X-ray and 
$\gamma$-ray bands. The resulting DCFs are shown in Figures \ref{DCF_RXB} to \ref{DCF_RGt}. 

The DCF reveals correlations/anticorrelations between two light curves, with a peak value of $\pm 1$ 
indicating a perfect correlation/anticorrelation, respectively. The time lag $\tau$ at which the peak
occurs, indicates a time lag between the variability patterns in the two light curves. In all cases,
the DCF results shown in Figures \ref{DCF_RXB} -- \ref{DCF_RGt} confirm the correlation and time lag
results apparent from Figures \ref{B_lightcurves} -- \ref{t_lightcurves} and Tables \ref{B_peaks}
-- \ref{t_peaks}. 

\begin{table}
\centering
\begin{tabular}{ccccc}
\hline
& $F_{1}$ & $ \sigma$ [s] & $\tau_{pk}$ [s] & Fig. \\
\hline
$R-X: B$  & $1.08$ & $(4.97 \pm 0.48) \times 10^{4}$ & $(-2.37 \pm 4.84) \times 10^{3}$ & \ref{DCF_RXB} \\
$R-\gamma: B$ & $-1.04$ & $(5.96 \pm 0.57) \times 10^{4}$ & $(5.96 \pm 0.57) \times 10^{4}$ & \ref{DCF_RGB} \\
$R-X: L_{inj}$  & $1.06$ & $(5.62 \pm 0.58) \times 10^{4}$ & $(1.94 \pm 0.51) \times 10^{4}$ & \ref{DCF_RXL} \\
$R-\gamma: L_{inj}$ & $1.11$ & $(4.97 \pm 0.49) \times 10^{4}$ & $(4.96 \pm 4.92) \times 10^{3}$ & \ref{DCF_RGL} \\
$R-X: t_{acc}$ & $-1.05$ & $(5.97 \pm 0.56) \times 10^{4}$ & $(3.22 \pm 0.54) \times 10^{4}$ & \ref{DCF_RXt} \\
$R-\gamma: t_{acc}$ & $1.13$ & $(5.39 \pm 0.49) \times 10^{4}$ & $(8.02 \pm 5.16) \times 10^{3}$ & \ref{DCF_RGt} \\
\hline
\end{tabular} \\ 
\caption{Best-fit DCF correlation strengths and time lags from the Gaussian fits to the discrete correlation 
functions.}
\label{Gaussian_parameters}
\end{table}

For a more rigorous analysis of the time lags and their errors, we performed a Gaussian fit of the form

\begin{equation}
DCF(\tau) = F_{1} \cdot e^{-(\tau - \tau_{pk})^{2}/2\sigma^{2}}
\label{DCF_Gaussian}
\end{equation}

\noindent to each of the discrete correlation functions around the peaks/troughs of the DCFs. This was
done by rigorous $\chi^{2}$ minimization. The minimization produces the best fit parameters for the 
normalization, $F_{1}$, the location of the peak time delay, $\tau_{pk}$, and the standard deviation 
of the peak, $\sigma$. The best fit parameters are listed in table \ref{Gaussian_parameters}. Within 
error bars, the time lags determined from the DCFs agree well with those extracted from inspection of 
the light curves (Tables \ref{B_peaks} -- \ref{t_peaks}).

\section{\label{results}Results and Discussion}

In this paper, we describe the development of a time-dependent model for the broadband emission
from blazars, incorporating Fermi II acceleration in the time evolution of the electron distribution,
as well as internal (SSC) and external (EC) target photon fields for Compton scattering, using the full
Klein-Nishina cross section. The use of a power law distribution for the electron injection spectrum is 
motivated by the physical picture of Fermi I acceleration providing the effective injection mechanism.
We consider two external radiation fields, namely the accretion disk and a second radiation field that
is approximated as being isotropic in the AGN rest frame, representative of either the BLR or of IR
emission from a dust torus. The code was used for a generic study of the influence of a diffusive
acceleration process on the equilibrium electron distribution and SEDs from blazars, and of the
multiwavelength radiative signatures of fluctuations of individual model parameters, including the
Fermi-II acceleration efficiency. The choice of baseline parameters was guided by a fit to the
time-averaged SED of the FSRQ 3C273, with our model simulation in a steady-state. Our study is
therefore representative of features expected for FSRQs or other low-frequency peaked blazars. 
We then investigated the potentially observable signatures of flaring activity caused by short-term 
fluctuations of (a) the magnetic field, (b) the electron injection luminosity, and (c) the acceleration 
time scale. A discrete correlation function analysis was performed on the light curves simulated
for the different flaring scenarios, at radio, optical, X-ray, and $\gamma$-ray frequencies to
quantify the predicted strengths of cross-band correlations and associated time lags. 

We found that magnetic field fluctuations lead to correlated radio, optical, and X-ray flaring, 
but an anti-correlation of these three bands with the $\gamma$-ray emission, with a time lag 
of up to several hours, due to increased synchrotron and SSC cooling of relativistic electrons. 
Flaring activity caused by fluctuations of the injection luminosity lead to correlated variability
in all wavelength ranges, with a time lag of a few hours between the optical and X-ray flares, associated
with the cooling time scale of electrons to reach low energies contributing to SSC X-ray emission. 
In this scenario, the radio emission shows a delayed drop in flux, due to the increase of the 
synchrotron self absorption caused by the increased number of low-energy electrons. A temporary
shortening of the acceleration time scale intensified both the synchrotron and Compton emission and
leads to a shift of both components to higher frequencies due to the acceleration of electrons to 
higher energies. Apart from correlated flaring activity at optical and $\gamma$-ray frequencies,
this has interesting consequences at X-rays and radio wavelengths: The shift of the SSC emission
to higher frequencies leads to a decrease of the (SSC-dominated) X-ray flux and therefore an anti-correlation
between the optical/$\gamma$-ray and X-ray fluxes with time delays of a few hours. At the same time,
the more efficient electron acceleration reduces the number of low-energy electrons responsible for
synchrotron self-absorption at radio wavelengths and therefore leads to a radio flare correlated
with the optical/$\gamma$-ray activity. 

Recent multi-wavelength observations of FSRQs have shown correlations between different 
wavelength bands that can be attributed to flares simulated in this paper. Multi-wavelength observations
of the FSRQ 3C 454.3 from August-December 2008 have shown pronounced flaring activity in the IR, UV,
x-ray and $\gamma$-ray bands with correlations for all bands except the x-rays \citep{Bonning09}. 
These correlations are consistent with a model in which a change in the injection luminosity of 
higher energy electrons takes place and interacts with external photons, causing the flaring 
observed in the $\gamma$-rays \citep{Bonning09}. The much longer cooling time of the low-energy
electrons responsible for the X-ray emission leads to much delayed variability, on much longer
time scales compared to the optical and $\gamma$-ray bands, which might be washed out by super-imposed
longer-term variability. Correlated multi-wavelength campaigns have also been done on the FSRQ 3C 273 
that reveal a correlation between the IR and x-ray bands, with time lags on the order of a few hours 
\citep{McHardy07}. This is consistent with the results presented here and supports the notion that
the X-ray emission is dominated by synchrotron self Compton radiation \citep{McHardy07}. 

We point out that our ad-hoc choice of Gaussian perturbations to key parameters only serves 
to study generic features of such changes. As has become obvious, the salient predictions 
concerning correlations/anti-correlations and time lags result from the microphysical processes 
of electron acceleration and cooling and are only weakly dependent on the exact time profile 
of the perturbation. 

\section*{Acknowledgments}

We thank the anonymous referee for a quick review and a constructive report which helped
to improve the manuscript. This work was funded by NASA through Astrophysics Data Analysis 
Program (ADAP) grant NNX12AE43G. The work of M.B. is supported through the South African 
Research Chair Initiative (SARChI) of the National Research Foundation and the Department 
of Science and Technology of South Africa, under SARChI Chair grant No. 64789.




\end{document}